\pdfoutput=1
\documentclass[a4paper,11pt]{article}
\usepackage[T1]{fontenc}
\usepackage[english]{babel}
\usepackage{amsmath}
\usepackage{libertine}
\usepackage[libertine]{newtxmath}
\usepackage{microtype}
\usepackage{xcolor}
\definecolor{linkcol}{rgb}{0, 0, 0.5}
\usepackage[top=25truemm,bottom=25truemm,
left=25truemm,right=25truemm]{geometry}
\usepackage{etoolbox}
\AtBeginEnvironment{thebibliography}{\small}
\usepackage{setspace}
\usepackage{cite}
\usepackage{titlesec}
\titleformat{\section}
{\large\bfseries\sffamily}{\thesection.}{0.5em}{}
\titleformat{\subsection}
{\normalsize\bfseries\sffamily}{\thesubsection.}{0.5em}{}
\usepackage{authblk}

\usepackage[pdfencoding=auto,
colorlinks,allcolors=linkcol,breaklinks=true]{hyperref}
\newcommand{\e}{\mathrm{e}}
\renewcommand{\d}{\mathrm{d}}
\renewcommand{\i}{\mathrm{i}}
\DeclareMathOperator{\sign}{sign}
\DeclareMathOperator{\Ai}{Ai}
\DeclareMathOperator{\Bi}{Bi}
\allowdisplaybreaks[2]

\title{\textbf{Unimodular quantum cosmology in the connection
    representation: A minimal model}}
\author{Shinji Yamashita\footnote{shinji0yamashita@gmail.com}}
\affil{National Institute of Technology, Niihama College, Ehime, Japan}
\date{\vspace{-5ex}}

\begin{document}
\maketitle

\begin{abstract}
  We present a quantization of unimodular gravity in the connection
  representation for a homogeneous, isotropic, and spatially flat
  cosmological model without matter.  In this model, the wave function
  is governed by a Schr\"odinger-type equation derived from a reduced
  phase space approach.  Our analysis suggests that, within this
  minimal setting, the regularity of the operators and the
  self-adjointness of the Hamiltonian operator are incompatible with a
  negative cosmological constant.  For a positive cosmological
  constant, the wave functions vanish at zero spatial volume.  This
  behavior emerges as a consequence of enforcing the unimodular
  condition at the quantum level.  Semiclassical fluctuations of the
  geometry are evaluated and discussed in relation to the cosmological
  constant problem.
\end{abstract}

\section{Introduction}

In canonical approaches to general relativity (GR), the dynamics is
governed by a set of constraints~\cite{Cianfrani2014}.  At the quantum
level, the Hamiltonian does not generate time evolution.  Rather, it
acts as a first-class constraint that annihilates physical states.
This structure obscures the description of quantum dynamics, giving
rise to the so-called problem of time.

Unimodular gravity is a conservative modification of GR.  In this
theory, the spacetime volume density is fixed, while the classical
Einstein field equations remain unchanged.  Although this modification
may appear minor, it leads to a distinctive feature at the canonical
level: namely, the Hamiltonian is no longer a constraint.  As a
result, the physical states obey a Schr\"odinger-type equation that
governs their time evolution~\cite{Unruh1989,Unruh1989a}.

In addition, unimodular gravity provides an alternative viewpoint on
the cosmological constant~\cite{Unruh1989,Unruh1989a,Smolin2009}.  At
the classical level, the cosmological constant emerges as an
integration constant rather than as a fixed parameter of the theory.
At the quantum level, it is naturally related to the eigenvalues of
the Hamiltonian, and the wave function is expressed as a superposition
of eigenstates.

Unimodular gravity can be formulated in two ways.  In the original
formulation of unimodular gravity, the unimodular condition is imposed
directly by fixing the spacetime volume
density~\cite{Bij1982,Kluson2015}.  In this case, general covariance
is restricted to volume-preserving diffeomorphisms.  Another
formulation is covariant unimodular gravity proposed by Henneaux and
Teitelboim~\cite{Henneaux1989}, in which a densitized vector field is
introduced to ensure full diffeomorphism invariance.

This paper investigates the properties of unimodular quantum gravity
in the connection representation.  The fundamental variables of this
representation are the Ashtekar-Barbero connection and a densitized
triad~\cite{Ashtekar1987,Barbero1995,Ashtekar2004}.  While the
connection representation is often regarded as the starting point of
loop quantum gravity, it remains of interest even without invoking
loop quantization.  A prominent exact solution in the connection
representation is the Chern-Simons-Kodama
state~\cite{Kodama1990,Smolin2002}.  In covariant unimodular gravity,
the Chern-Simons-Kodama state can be consistently
extended~\cite{Smolin2011,Yamashita2021,Alexandre2023}, whereas this
is not the case in the original unimodular
gravity~\cite{Yamashita2020}.  Despite this difference, we adopt the
original unimodular gravity and apply it to a minimal cosmological
model in which spacetime is homogeneous, isotropic, and spatially flat
without matter.  This minimal setting allows us to isolate specific
effects arising from the unimodular condition.

The implementation of the unimodular condition depends on the choice
of fundamental variables.  In the usual metric formulation, the
unimodular condition is imposed on the determinant of the spacetime
metric, whereas in the connection formulation, it is imposed on the
determinant of the tetrad.  This difference leads to subtle
differences in the canonical structure which may become nontrivial at
the quantum level.

This paper is organized as follows.  In
section~\ref{sec:classical_theory}, we perform the Hamiltonian
analysis of the original unimodular gravity in the simplest
cosmological model.  The phase space is reduced by solving the
second-class constraints.  After the reduction, we obtain the reduced
Hamiltonian and the corresponding classical solution.  In
section~\ref{sec:quantum_theory}, we carry out the canonical
quantization and obtain exact solutions.  Additionally, we consider a
semiclassical description via the Wentzel-Kramers-Brillouin (WKB)
approximation.  In particular, we examine the fluctuation of geometric
quantities around the classical trajectory.  In
section~\ref{sec:summary}, we summarize and discuss how unimodularity
modifies the structure of canonical quantum cosmology.

We employ the following notation.  The spacetime metric has signature
$(-,+,+,+)$.  Capital Latin letters $I,J,\cdots \in \{0,1,2,3\}$
denote Lorentz indices.  Greek letters
$\mu,\nu,\cdots \in \{\tau,1,2,3\}$ denote spacetime indices, with
$\tau$ labeling the time coordinate.  Lowercase Latin letters
$i,j,\dots \in \{1,2,3\}$ and $a,b,\cdots \in \{1,2,3\}$ denote
internal and spatial indices, respectively.  We use units in which the
speed of light is unity.

\section{Classical theory}
\label{sec:classical_theory}

\subsection{Constraints and Hamiltonian}

The fundamental variables of our theory are a connection $A_{a}^{i}$
and a densitized triad $E_{i}^{a}$, defined by
\begin{align}
  A_{a}^{i}
  &= -\frac{1}{2}\epsilon^{i}{}_{jk}\omega_{a}^{jk}
    -\beta\omega_{a}^{0i},
    \label{eq:1}\\
  E_{i}^{a}
  &= \left(\det e_{a}^{i}\right) e_{i}^{a},
    \label{eq:2}
\end{align}
where $\omega_{\mu}^{IJ}$ is a spin connection compatible with a
tetrad $e_{I}^{\mu}$, $\beta$ is the real-valued Barbero-Immirzi
parameter, and $e_{i}^{a}$ is a triad.  We consider the following
action $S$, which is based on the Holst action~\cite{Holst1996}:
\begin{align}
  S = -\frac{1}{2k\beta}\int e^{I} \wedge e^{J} \wedge \left(
  R_{IJ} - \frac{\beta}{2}\epsilon_{IJKL}R^{KL}\right)
  -\frac{1}{24k}\int \Lambda \epsilon_{IJKL} e^{I}\wedge e^{J} \wedge
  e^{K} \wedge e^{L}
  +\frac{1}{k}\int \d^{4}x\ \Lambda \alpha,
  \label{eq:3}
\end{align}
where $k$ is equal to $8\pi$ times Newton's constant, $R^{IJ}$ is the
curvature of $\omega_{\mu}^{IJ}$, $\Lambda$ is a scalar field that
plays the role of the unimodular cosmological constant, and $\alpha$
is a nonvanishing fixed scalar density.  Varying the action with
respect to $\Lambda$ yields the unimodular condition
\begin{align}
  \det e_{\mu}^{I}+\alpha=0.
  \label{eq:4}
\end{align}
By performing a $3+1$ decomposition in the time gauge, which assigns
$e_{\tau}^{0}=-N, e_{\tau}^{i}=N^{a}e_{a}^{i}, e_{a}^{0}=0$ with the
lapse function $N$ and the shift vector $N^{a}$, we obtain the action
in the form
\begin{align}
  S=\frac{1}{k\beta} \int \d^{4}x \biggl[E_{i}^{a}\dot{A}_{a}^{i}
  -A_{\tau}^{i}G_{i}-N^{a}V_{a}-NC
  - \beta\Lambda
  (N\det e_{a}^{i} - \alpha)\biggr].
  \label{eq:5}
\end{align}
Here,
\begin{align}
  G_{i}
  &=- \left(\partial_{a}E_{i}^{a}
    + \epsilon_{ij}{}^{k}A_{a}^{j}E_{k}^{a}\right),
    \label{eq:6} \\
  V_{a}
  &=-E_{i}^{b}F_{ba}^{i}, \label{eq:7} \\
  C
  &=\frac{\beta}{2\det e_{a}^{i}} \epsilon^{ijk}E_{i}^{a}E_{j}^{b}
    \left[F_{abk} - (1+\beta^{2}) \epsilon_{klm}K_{a}^{l}K_{b}^{m}
    \right],
    \label{eq:8}
\end{align}
$F_{ab}^{i}$ is the curvature of $A_{a}^{i}$, and
$K_{a}^{i}=-\omega_{a}^{0i}=K_{ab}e_{j}^{b}\delta^{ij}$ is related to
the extrinsic curvature $K_{ab}$.

We now consider a cosmological model in which the spacetime is
homogeneous, isotropic, and spatially flat without matter.  The line
element is assumed to take the form
\begin{align}
  \d s^{2} = -N^{2}\d \tau^{2}
  + a^{2}(\tau)e^{(o)}{}_{a}^{i}e^{(o)}{}_{b}^{j}\delta_{ij}
  \d x^{a}\d x^{b},
  \label{eq:9}
\end{align}
where $e^{(o)}{}_{a}^{i}$ is a fiducial cotriad, and $a(\tau)$ denotes
a scale factor.  We assume that the spatial manifold is topologically
$\mathbb{R}^{3}$.  Since the spatial manifold is non-compact, we
restrict all spatial integrations to a fiducial cell of
three-volume~\cite{Bojowald2011}
\begin{align}
  V_{o}=\int \d^{3}x\ \left|\det e^{(o)}{}_{a}^{i}\right|.
  \label{eq:10}
\end{align}
The connection and densitized triad consistent with the symmetry
reduction take the form
\begin{align}
  A_{a}^{i}
  &=c(\tau) V_{o}^{-\frac{1}{3}} e^{(o)}{}_{a}^{i},
    \label{eq:11} \\
  E_{i}^{a}
  &=p(\tau)V_{o}^{-\frac{2}{3}} \left|\det e^{(o)}{}_{a}^{i}\right|
    e^{(o)}{}_{i}^{a}.
    \label{eq:12}
\end{align}
The dynamical variables $c(\tau)$ and $p(\tau)$ are related to the
scale factor by
\begin{align}
  c(\tau)
  &=V_{o}^{\frac{1}{3}} \frac{\beta}{N}\partial_{\tau}a(\tau),
    \label{eq:13} \\
  p(\tau)&=V_{o}^{\frac{2}{3}} s a^{2}(\tau),
           \label{eq:14}
\end{align}
where
\begin{align}
  s = \sign \left(\det e_{a}^{i}\right).
  \label{eq:15}
\end{align}
Substituting the symmetry-reduced expressions \eqref{eq:11} and
\eqref{eq:12} into the $3+1$ decomposed action \eqref{eq:5}, and
integrating over the spatial coordinates, we obtain
\begin{align}
  S = \int \d \tau\ L
  =\int \d \tau \left[\frac{3}{k\beta} p \dot{c}
  +\frac{3sN}{k\beta^{2}} \left|p\right|^{\frac{1}{2}} c^{2}
  -\frac{\Lambda}{k} \left(sN \left|p\right|^{\frac{3}{2}}
  -\mathcal{E}\right) \right],
  \label{eq:16}
\end{align}
where $L$ is the reduced Lagrangian, and the quantity
\begin{align}
  \mathcal{E} = \int \d^{3}x\ \alpha
  \label{eq:17}
\end{align}
is a nonvanishing constant determined by the scalar density $\alpha$.
Note that $G_{i}$ \eqref{eq:6} and $V_{a}$ \eqref{eq:7} vanish
automatically due to homogeneity and isotropy.  The configuration
variables of this model are $\left(c, N, \Lambda\right)$, with the
corresponding canonical momenta
$\left(3 \left(k\beta\right)^{-1}p, p_{N}, p_{\Lambda}\right)$ defined
by
\begin{align}
  \frac{\partial L}{\partial \dot{c}}
  &= \frac{3}{k\beta}p,
    \label{eq:18} \\
  \frac{\partial L}{\partial \dot{N}}
  &= p_{N} = 0,
    \label{eq:19} \\
  \frac{\partial L}{\partial \dot{\Lambda}}
  &= p_{\Lambda} = 0.
    \label{eq:20}
\end{align}
Equations \eqref{eq:19} and \eqref{eq:20} yield the following primary
constraints
\begin{align}
  p_{N} \approx 0,
  \label{eq:21}\\
  p_{\Lambda} \approx 0,
  \label{eq:22}
\end{align}
where the symbol $\approx$ denotes weak equality, indicating that
the equation holds on the constraint surface.  The total Hamiltonian
$H_{\text{T}}$ is defined as the sum of the ordinary canonical
Hamiltonian and the primary constraints with the Lagrange multipliers
$v_{N}$ and $v_{\Lambda}$:
\begin{align}
  H_{\text{T}}=-\frac{3sN}{k\beta^{2}}
  \left|p\right|^{\frac{1}{2}} c^{2}
  +\frac{\Lambda}{k} \left(sN \left|p\right|^{\frac{3}{2}}
  -\mathcal{E}\right) + v_{N}p_{N} + v_{\Lambda}p_{\Lambda}.
  \label{eq:23}
\end{align}
Each constraint must satisfy the stability condition, which ensures
that the constraint is preserved under time evolution.  The stability
conditions for the primary constraints \eqref{eq:21} and \eqref{eq:22}
yield
\begin{align}
  \left\{p_{N}, H_{\text{T}}\right\}
  &=-\Phi \approx 0, \label{eq:24}
  \\
  \left\{p_{\Lambda}, H_{\text{T}}\right\}
  &= -\Theta \approx 0, \label{eq:25}
\end{align}
which lead to the secondary constraints
\begin{align}
  \Phi
  &=-\frac{3s}{k\beta^{2}} \left|p\right|^{\frac{1}{2}}c^{2}
    +\frac{s\Lambda}{k} \left|p\right|^{\frac{3}{2}} \approx 0,
    \label{eq:26} \\
  \Theta
  &=\frac{1}{k} \left(sN |p|^{\frac{3}{2}}
    -\mathcal{E}\right) \approx 0.
    \label{eq:27}
\end{align}
Constraints $\Phi\approx 0$ and $\Theta\approx 0$ correspond to the
Hamiltonian constraint and the unimodular constraint, respectively.
The stability conditions for $\Phi$ and $\Theta$ become
\begin{align}
  \left\{\Phi, H_{\text{T}}\right\}
  &=\frac{s}{k} \left|p\right|^{\frac{3}{2}}v_{\Lambda}
    \approx 0, \label{eq:28} \\
  \left\{\Theta, H_{\text{T}}\right\}
  &\approx \frac{3sN^{2}}{k\beta} |p|c
    +\frac{s}{k} |p|^{\frac{3}{2}} v_{N} \approx 0.
    \label{eq:29}
\end{align}
Therefore, these conditions determine $v_{\Lambda}$ and $v_{N}$ as
\begin{align}
  v_{\Lambda} = 0, \quad
  v_{N} = -\frac{3N^{2}}{\beta} |p|^{-\frac{1}{2}}c,
  \label{eq:30}
\end{align}
and no further constraints arise.  The pairs of the constraints with
weakly nonvanishing Poisson brackets are
\begin{align}
  \left\{p_{N}, \Theta\right\}
  &= -\frac{s}{k} |p|^{\frac{3}{2}} \not\approx 0,
    \label{eq:31} \\
  \left\{p_{\Lambda}, \Phi\right\}
  &=-\frac{s}{k}|p|^{\frac{3}{2}} \not\approx 0,
    \label{eq:32} \\
  \left\{\Phi, \Theta\right\}
  &= -\frac{3sN}{k\beta}|p|c \not\approx 0.
    \label{eq:33}
\end{align}
These results imply that all constraints
$\left(p_{N},p_{\Lambda},\Phi,\Theta\right)$ are second class.  At
this stage, there are two different ways of dealing with the
second-class constraints.  The first is to keep the second-class
constraints, and employ the Dirac bracket formalism.  The second is to
solve the second-class constraints explicitly, reduce the phase space
by expressing the dependent variables in terms of the independent
ones, and carry out the analysis using the Poisson
brackets~\cite{Henneaux1992,Matschull1996}.  In the present work, we
adopt the latter approach.  This choice is motivated by subsequent
quantization, since the Dirac bracket generally leads to a complicated
operator algebra at the quantum level.

From the primary constraints \eqref{eq:21} and \eqref{eq:22}, the
momenta $p_{N}$ and $p_{\Lambda}$ are fixed as
\begin{align}
  p_{N}=p_{\Lambda}=0.
  \label{eq:34}
\end{align}
From the secondary constraints \eqref{eq:26} and \eqref{eq:27}, the
dependent variables $\Lambda$ and $N$ are expressed in terms of the
independent variables $c$ and $p$ as
\begin{align}
  \Lambda
  &= \frac{3}{\beta^{2}} |p|^{-1}c^{2},
    \label{eq:35} \\
  N
  &= s\mathcal{E} |p|^{-\frac{3}{2}}.
    \label{eq:36}
\end{align}
Substituting Eqs. \eqref{eq:34}--\eqref{eq:36} into the total
Hamiltonian \eqref{eq:23}, we obtain the reduced Hamiltonian $H$
expressed in terms of the independent variables $c$ and $p$ as
\begin{align}
  H = -\frac{3\mathcal{E}}{k\beta^{2}} |p|^{-1}c^{2}.
  \label{eq:37}
\end{align}
Unlike general relativity, the Hamiltonian $H$ is not a
constraint, and it can be expressed in terms of $\Lambda$
\eqref{eq:35} as
\begin{align}
  H=-\frac{\mathcal{E}}{k}\Lambda.
  \label{eq:38}
\end{align}
Since $\left\{\Lambda, H\right\}=0$, the unimodular cosmological
constant $\Lambda$ is conserved under time evolution.

\subsection{Classical solution}

Before proceeding to quantization, we examine the classical dynamics.
The equations of motion for $c$ and $p$ are
\begin{align}
  \dot{c}
  &= \left\{c,H\right\}
    = \frac{s\mathcal{E}}{\beta} \frac{c^{2}}{|p|^{2}},
    \label{eq:39} \\
  \dot{p}
  &= \left\{p,H\right\}
    = \frac{2\mathcal{E}}{\beta} \frac{c}{|p|}.
    \label{eq:40}
\end{align}
Using equations \eqref{eq:13} and \eqref{eq:14} together with the above
equations of motion, we obtain the classical solution for the scale
factor $a(\tau)$ as
\begin{align}
  a(\tau) = \left(C_{1}\tau + C_{2}\right)^{\frac{1}{3}},
  \label{eq:41}
\end{align}
where $C_{1}$ and $C_{2}$ are constants.
It is convenient to introduce the comoving time $t$ defined by
\begin{align}
  \d t = N \d\tau= \frac{s\mathcal{E}}{V_{o}}
  \frac{1}{C_{1}\tau + C_{2}} \d \tau.
  \label{eq:42}
\end{align}
The cube of the scale factor can be written in the exponential form
\begin{align}
  a^{3}(t) = C_{3} \exp \left[\frac{C_{1}V_{o}}{s\mathcal{E}}t\right],
  \label{eq:43}
\end{align}
with a constant $C_{3}$.

\section{Quantum theory}
\label{sec:quantum_theory}

\subsection{Exact solutions}

Following the canonical quantization procedure, the classical Poisson
bracket relation $\left\{c, 3(k\beta)^{-1}p\right\}=1$ is replaced
by the corresponding quantum commutation relation
\begin{align}
  \left[\hat{c}, \frac{3}{k\beta}\hat{p}\right] = \i\hbar.
  \label{eq:44}
\end{align}
In the $p$-representation, the fundamental operators act on a
wave function $\Psi$ as
\begin{align}
  \hat{c}\Psi
  &=\frac{\i\hbar k\beta}{3}\frac{\partial \Psi}{\partial p},
    \label{eq:45} \\
  \hat{p}\Psi
  &= p\Psi.
    \label{eq:46}
\end{align}
Since the Hamiltonian in unimodular gravity is not a constraint, the
quantum dynamics is governed by a Schr\"odinger-type equation
\begin{align}
  \i\hbar\partial_{\tau} \Psi(\tau,p) = \hat{H}\Psi(\tau,p).
  \label{eq:47}
\end{align}
The Hamiltonian operator $\hat{H}$ involves products of the
noncommuting operators $\hat{c}$ and $\hat{p}$, which lead to an
operator-ordering ambiguity.  In the present work, we choose a
symmetric factor ordering.  With this choice, the Hamiltonian operator
$\hat{H}$ is defined as
\begin{align}
  \hat{H} = -\frac{3\mathcal{E}}{k\beta^{2}}
  |\hat{p}|^{-\frac{1}{2}} \hat{c}^{2} |\hat{p}|^{-\frac{1}{2}}.
  \label{eq:48}
\end{align}
We consider a stationary state $\Psi_{\omega}(\tau,p)$, which
satisfies the eigenvalue equation
\begin{align}
  \i\hbar \partial_{\tau}\Psi_{\omega}(\tau,p)
  = \hat{H}\Psi_{\omega}(\tau,p)=\hbar \omega \Psi_{\omega}(\tau,p),
  \label{eq:49}
\end{align}
where $\hbar\omega$ denotes the eigenvalue of the Hamiltonian.
To solve \eqref{eq:49}, we assume that $\Psi_{\omega}$ has the form
\begin{align}
  \Psi_{\omega}(\tau,p)
  =\e^{-\i\omega\tau} |p|^{\frac{1}{2}}\chi_{\omega}(p).
  \label{eq:50}
\end{align}
Substituting \eqref{eq:50} into \eqref{eq:49}, we obtain the following
equation for $\chi_{\omega}(p)$:
\begin{align}
  \frac{\d^{2}\chi_{\omega}}{\d p^{2}}-A|p|\chi_{\omega}=0,
  \label{eq:51}
\end{align}
where the constant $A$ is given by
\begin{align}
  A=\frac{3\omega}{\mathcal{E}\hbar k}.
  \label{eq:52}
\end{align}
Using \eqref{eq:52} together with \eqref{eq:38}, $A$ can be expressed
in terms of $\Lambda$ as
\begin{align}
  A=-\frac{3\Lambda}{\hbar^{2}k^{2}}.
  \label{eq:53}
\end{align}
Therefore the sign of $A$ is determined by the sign of $\Lambda$.

For $A\neq 0$, the solutions of the differential equation
\eqref{eq:51} are expressed in terms of the Airy functions $\Ai$ and
$\Bi$.  For $A>0\ (\Lambda < 0)$, the eigenstate takes the form
\begin{align}
  \Psi_{\omega}^{(A+)}(\tau,p)
  =\e^{-\i\omega\tau} |p|^{\frac{1}{2}}
  \left[C_{1}\Ai \left(A^{\frac{1}{3}}|p|\right)
  +C_{2}\Bi \left(A^{\frac{1}{3}}|p|\right) \right],
  \label{eq:54}
\end{align}
with constants $C_{1}$ and $C_{2}$.  Since the second term grows
exponentially, it should be discarded by setting $C_{2}=0$.  This
non-oscillatory solution can be interpreted as corresponding to a
classically forbidden region, in analogy with ordinary quantum
mechanics.

For $A<0\ (\Lambda>0)$, the eigenstate is given by
\begin{align}
  \Psi_{\omega}^{(A-)}(\tau,p)
  =\e^{-\i\omega\tau} |p|^{\frac{1}{2}}
  \left[D_{1}\Ai \left(-|A|^{\frac{1}{3}}|p|\right)
  +D_{2}\Bi \left(-|A|^{\frac{1}{3}}|p|\right) \right],
  \label{eq:55}
\end{align}
with constants $D_{1}$ and $D_{2}$.  This state is oscillatory and
grows as $|p|^{1/4}$ for large $|p|$.  It can be interpreted
as corresponding to a classically allowed region.

We discuss several aspects of their interpretations and physical
implications.  An important difference from quantum cosmology in terms
of the scale factor $a$ is that $p=V_{o}^{2/3}sa^{2}$ can take all
real values.  Thus, the wave function is defined
across the point $p=0$.  Since the derivative of
$\Psi_{\omega}^{(A-)}$ with respect to $p$ is
\begin{align}
  \frac{\partial \Psi_{\omega}^{(A-)}}{\partial p}
  =\e^{-\i\omega\tau} \left[
  \frac{s}{2}|p|^{-\frac{1}{2}}\chi_{\omega}
  +|p|^{\frac{1}{2}}\frac{\d \chi_{\omega}}{\d p}\right],
  \label{eq:56}
\end{align}
the eigenstate $\Psi_{\omega}^{(A-)}$ is not regular at $p=0$ due to
the presence of the first term in \eqref{eq:56}.  To retain the
regularity of the wave function, we impose
\begin{align}
  \chi_{\omega}(p=0)
  =D_{1}\Ai(0) + D_{2}\Bi(0) = 0,
  \label{eq:57}
\end{align}
which leads to
\begin{align}
  D_{1} = -\frac{\Bi(0)}{\Ai(0)}D_{2} = -\sqrt{3}D_{2}.
  \label{eq:58}
\end{align}
This coefficient ratio is uniquely determined by requiring the
fundamental operators to be well-defined.  First, this ratio ensures
the regularity of the operator $\hat{c}$.  Since $\hat{c}$ involves a
derivative with respect to $p$, the wave function should be regular at
$p=0$ to avoid divergence of $\hat{c}\Psi$.  Furthermore, this
coefficient ratio also ensures the self-adjointness of the Hamiltonian
operator.  More specifically, with the inner product
$\left\langle\Phi\middle|\Psi\right\rangle =\int_{-\infty}^{+\infty} \d p\ \Phi^{*}\Psi$,
the self-adjointness of $\hat{H}$ requires
\begin{align}
  \left\langle
  \hat{H}\Phi_{\omega'}\middle|\Psi_{\omega}\right\rangle
  - \left\langle\Phi_{\omega'}\middle|\hat{H}\Psi_{\omega}\right
  \rangle
  &\sim \int_{-\infty}^{+\infty} \d p\ \frac{\partial}{\partial p}
    \left[
    |p|^{-1} \left(\frac{\partial \Phi_{\omega'}^{*}}{\partial p}
    \right)\Psi_{\omega} - |p|^{-1}\Phi_{\omega'}^{*}
    \frac{\partial \Psi_{\omega}}{\partial p} \right]
    \notag \\
  &\sim \lim_{\epsilon\to 0} \biggl(
    \left[\frac{\d \chi_{\omega'}}{\d p} \chi_{\omega}
    -\chi_{\omega'} \frac{\d \chi_{\omega}}{\d p}
    \right]_{-\infty}^{-\epsilon}
    +\left[\frac{\d \chi_{\omega'}}{\d p} \chi_{\omega}
    -\chi_{\omega'} \frac{\d \chi_{\omega}}{\d p}
    \right]_{+\epsilon}^{\infty}\biggr) = 0.
    \label{eq:59}
\end{align}
The boundary contribution at $p\to \pm\infty$ can be eliminated by
constructing a wave packet as a superposition of the oscillatory
eigenstates over $\omega$.  On the other hand, since the divergence of
the derivative at $p=0$ is common to all eigenstates
$\Psi_{\omega}^{(A-)}$, it cannot be eliminated by any superposition.
The coefficient ratio \eqref{eq:58} removes this singularity and
guarantees the self-adjointness condition \eqref{eq:59}.

When applying the same coefficient ratio \eqref{eq:58} to
$\Psi_{\omega}^{(A+)}$ \eqref{eq:54}, we find that the possible
coefficients are $C_{1}=-\sqrt{3} C_{2}=0$, which yield a trivial
state $\Psi_{\omega}^{(A+)}=0$.  Therefore, for $A>0\ (\Lambda<0)$,
this theory cannot simultaneously ensure the suppression of the
exponential growth of the wave function at large $|p|$ and the
self-adjointness of fundamental operators such as $\hat{c}$ and
$\hat{H}$.  This inconsistency suggests that, at least in the minimal
setting, a negative cosmological constant cannot be consistently
realized even at the quantum level.  This feature corresponds to the
classical impossibility for the Friedmann equation
$(\dot{a}/a)^{2} = \Lambda/3 <0$.

For $A=\Lambda=\omega=0$, the equation \eqref{eq:49} yields a
non-oscillatory solution $\Psi^{(A0)} \sim |p|^{\frac{3}{2}}$.  Unlike
the oscillatory Airy functions, the monotonic growth of $\Psi^{(A0)}$
cannot be canceled out through the interference of the different
modes in a wave packet.  Therefore, $\Psi^{(A0)}$ prevents the
construction of the normalizable wave packet and will be discarded.

When the sign of $p$ changes from negative to positive, the
consistency with unimodularity requires that both the factors
$s=\sign \left(\det e_{a}^{i}\right)$ and $\sign(N)$ flip
simultaneously, because the sign of the constant $\mathcal{E}$
\eqref{eq:17} is given by
\begin{align}
  \sign(\mathcal{E}) = s\sign(N).
  \label{eq:60}
\end{align}
This implies that the theory is invariant under the simultaneous
flipping of the spatial and temporal orientations.

Finally, for $A<0\ (\Lambda > 0)$, the eigenstates vanish at $p=0$.
The origin of this behavior can be traced to the unimodular
constraint.  The unimodular constraint \eqref{eq:27} requires
$N=s\mathcal{E}|p|^{-3/2}$ \eqref{eq:36}, which diverges at $p\to 0$.
Substituting this into the total Hamiltonian \eqref{eq:23}, we obtain
the reduced Hamiltonian \eqref{eq:37} with the singular factor
$|p|^{-1}$.  To solve the eigenvalue equation \eqref{eq:49} with this
singular operator, the eigenstates are required to have the factor
$|p|^{1/2}$, which naturally leads to vanishing wave functions at
$p=0$.  Therefore, this behavior can be understood as a direct
consequence of enforcing unimodularity.

\subsection{Semiclassical analysis for $A<0$}

We now perform the semiclassical analysis for $A<0\ (\Lambda > 0)$,
where the exact solution \eqref{eq:55} is oscillatory.  From the
eigenvalue equation \eqref{eq:49} for $A<0$, the WKB approximation
yields the state
\begin{align}
  \Psi_{\omega}^{\text{(WKB)}} (\tau,p)
  =\e^{\i\phi}
  =\exp \left[\i \left(-\omega\tau
  \pm \frac{2}{3}\sqrt{|A|} |p|^{\frac{3}{2}}\right)\right],
  \label{eq:61}
\end{align}
where the phase factor $\phi$, which is classically Hamilton's
principal function times $\hbar^{-1}$, is defined as
\begin{align}
  \phi= -\omega\tau \pm \frac{2}{3}\sqrt{|A|}|p|^{\frac{3}{2}}
  =-\omega\tau \pm \frac{2}{3}
  \left|\frac{3\omega}{\mathcal{E}\hbar k}\right|^{\frac{1}{2}}
  |p|^{\frac{3}{2}}.
  \label{eq:62}
\end{align}
This approximation is valid in the semiclassical regime where
$\sqrt{|A|} |p|^{3/2} \gg 1$.

To construct a physically relevant state, we consider a wave packet by
superposing the eigenstates with a distribution function $f(\omega)$:
\begin{align}
  \Psi(\tau,p)
  =\int \d \omega\ f(\omega)\Psi_{\omega}^{\text{(WKB)}}(\tau,p).
  \label{eq:63}
\end{align}
We assume that $f(\omega)$ is a Gaussian distribution which is sharply
peaked around some central value of $\omega$.  To find where
the wave packet \eqref{eq:63} is localized in $(\tau, p)$-space, or
equivalently, where the wave packet remains coherent, we seek the
region where the phase $\phi$ is stationary with respect to
$\omega$.  The stationary condition is given by
\begin{align}
  \frac{\partial \phi}{\partial \omega}
  =-\tau \pm \frac{\sign(\omega)}{3} \left|
  \frac{3}{\mathcal{E}\hbar k\omega} \right|^{\frac{1}{2}}
  |p|^{\frac{3}{2}} = 0.
  \label{eq:64}
\end{align}
Solving this equation for $|p|^{3/2}=a^{3}V_{o}$, we obtain
\begin{align}
  |p|^{\frac{3}{2}}=a^{3}V_{o}
  =\pm \sign(\omega)|3\mathcal{E}\hbar k\omega|^{\frac{1}{2}}\tau.
  \label{eq:65}
\end{align}
Note that the classical solution \eqref{eq:41} is consistent with
\eqref{eq:65}.

We examine the semiclassical fluctuation around this classical
trajectory.  Equation \eqref{eq:65} implies that the fluctuation in
$\omega$ leads to the fluctuation in $|p|^{3/2}$ which is given by
\begin{align}
  \delta |p|^{\frac{3}{2}}
  = \frac{|p|^{\frac{3}{2}}}{2\omega} \delta \omega.
  \label{eq:66}
\end{align}
Additionally, the unimodular condition implies that the fluctuation in
$|p|^{3/2}$ induces the fluctuation in the lapse $N$.  By
using \eqref{eq:36}, we have
\begin{align}
  \delta N
  =-N|p|^{-\frac{3}{2}} \delta |p|^{\frac{3}{2}}.
  \label{eq:67}
\end{align}
Combining \eqref{eq:66} and \eqref{eq:67}, we obtain the relationship
between the relative fluctuations:
\begin{align}
  \frac{\delta |p|^{\frac{3}{2}}}{|p|^{\frac{3}{2}}}
  =-\frac{\delta N}{N}=\frac{\delta \omega}{2\omega}=r,
  \label{eq:68}
\end{align}
where $r$ denotes the common value of the relative fluctuations.
Substituting \eqref{eq:65} into \eqref{eq:62}, the extremum value of
the phase $\phi_{\text{ex}}$ is expressed as
\begin{align}
  \phi_{\text{ex}} = \omega\tau.
  \label{eq:69}
\end{align}
The fluctuation in the phase $\phi$ around this extremum is given by
$\delta \phi_{\text{ex}}=\tau\delta\omega$.  The wave packet remains
coherent as long as the phase fluctuation $\delta \phi_{\text{ex}}$ is
of order unity.  When $\delta \phi_{\text{ex}}$ becomes much larger
than unity, the coherence of the wave packet is lost.  The coherence
time $\tau_{\text{coh}}$ is defined as the time at which the phase
fluctuation $\delta \phi_{\text{ex}}$ becomes of order unity, namely,
$\delta \phi_{\text{ex}}= \tau_{\text{coh}} \delta\omega \sim 1$.
This relation leads to $\delta \omega \sim 1/\tau_{\text{coh}}$, and
then
\begin{align}
  r = \frac{\delta\omega}{2\omega}
  \sim \frac{1}{\omega \tau_{\text{coh}}}.
  \label{eq:70}
\end{align}
Equations \eqref{eq:52} and \eqref{eq:53} imply
$\omega=-\mathcal{E}\Lambda/\hbar k$.  Additionally, under the
unimodular condition, the spacetime four-volume $V_{4}$ at the
coherence time $\tau_{\text{coh}}$ is expressed as
\begin{align}
  V_{4}=\int \d^{4}x\ \left|\det e_{\mu}^{I}\right|
  = \left|\mathcal{E}\right|\tau_{\text{coh}}.
  \label{eq:71}
\end{align}
Substituting these relations into \eqref{eq:70}, we obtain
\begin{align}
  r \sim \frac{\hbar k}{\Lambda V_{4}}.
  \label{eq:72}
\end{align}
It is worth emphasizing that the factor $\hbar k/(\Lambda V_{4})$ is
proportional to the inverse of Hamilton's principal function, namely,
the action evaluated on the classical trajectory.  In this sense, this
result is consistent with the standard WKB estimate of fluctuation.
The factor $\hbar k/(\Lambda V_{4})$ also appears in standard GR as a
measure of semiclassical fluctuations of the geometry.

In the present work, however, its origin is fundamentally different.
In standard GR, the cosmological constant is a fixed parameter,
whereas in unimodular gravity, it appears as an eigenvalue and a
physical state is described by a superposition over different values
of $\Lambda$.  The factor $\hbar k/(\Lambda V_{4})$ characterizes how
the coherence of the wave packet depends on the value of $\Lambda$
itself.

To evaluate this expression numerically and relate it to the
observable Universe, we assume that the coherence time
$\tau_{\text{coh}}$ is identified with the present age of the
Universe.  Under this assumption, the present Universe can be regarded
as a semiclassical coherent state.  Since we consider a Universe
dominated by a positive cosmological constant, the Hubble parameter
can be approximated as $\sqrt{\Lambda/3}$.  In such a de Sitter-like
phase, the characteristic four-volume is the Hubble four-volume
$V_{4}=9/\Lambda^{2}$.  The quantity $\hbar k$ is of the order of the
square of the Planck length, whereas the theoretical value of the
cosmological constant $\Lambda_{\text{theo}}$, estimated from vacuum
energy considerations, is proportional to the inverse square of the
Planck length.  Therefore, $\hbar k\sim \Lambda_{\text{theo}}^{-1}$.
Substituting these results yields
\begin{align}
  r \sim \frac{\Lambda}{\Lambda_{\text{theo}}} \sim 10^{-120},
  \label{eq:73}
\end{align}
where we have used the well-known inconsistency between the observed
cosmological constant $\Lambda$ and its theoretical value
$\Lambda_{\text{theo}}$.

Although the step from \eqref{eq:72} to \eqref{eq:73} is algebraically
identical to that in standard GR, its physical interpretation is
different.  In unimodular gravity, the physical state is described by
a superposition over different values of $\Lambda$.  From this
perspective, equation \eqref{eq:73} indicates that the smallness of
$\Lambda$ ensures the suppression of the spread of the wave packet,
and hence its semiclassical coherence.  This feature originates from
the unimodular structure, in which $\Lambda$ appears as an eigenvalue
rather than a fixed parameter of the theory.

\section{Summary and discussion}
\label{sec:summary}

In this paper, we have investigated the original formulation of
unimodular gravity in the simplest cosmological model.  The aim of
this work is to isolate the effects arising from the unimodular
condition and clarify how they modify the structure of canonical
quantum cosmology.

At the classical level, the unimodular cosmological constant $\Lambda$
\eqref{eq:35} and the lapse function $N$ \eqref{eq:36} are not treated
as independent canonical variables: instead, they are determined in
terms of $c$ and $p$ by employing the reduced phase space approach.

After carrying out the canonical quantization, we have defined a
Schr\"odinger-type equation by using the reduced Hamiltonian operator
\eqref{eq:48}.  Within this framework, $\Lambda$ is related to the
eigenvalues of the Hamiltonian operator.  We have derived exact
eigenstates of the Hamiltonian operator in terms of the Airy
functions.

The structure of eigenstates of the Hamiltonian depends on the sign of
the cosmological constant.  For a positive cosmological constant, the
eigenstates \eqref{eq:55} are oscillatory and can be used to construct
normalizable wave packets.  In contrast, for a negative cosmological
constant, the requirement of square-integrability at large $|p|$ is
incompatible with the regularity of the operators and the
self-adjointness of the Hamiltonian at $p=0$.  As a result, at least
in our minimal cosmological setting, a negative cosmological constant
is not compatible.

A similar conclusion has been obtained in other quantization schemes.
Specifically, Refs.~\cite{Amadei2019a,Amadei2019b,Amadei2023}
demonstrate that the original unimodular gravity formulated in terms
of $v\propto a^{3}$ and $b\propto \dot{v}$ can be mapped to a
non-relativistic free particle, where the positivity of the
Hamiltonian spectrum naturally excludes a negative cosmological
constant.  While our model reaches this conclusion through the
requirements of the regularity of $\hat{c}$ and the self-adjointness
of the Hamiltonian, both frameworks exclude a negative cosmological
constant in the vacuum case.  Furthermore, in trace-free Einstein
gravity~\cite{Montesinos2025}, which shares some structural features
with unimodular gravity but is a distinct theory, a similar feature is
observed.  In this framework, the cosmological constant has a positive
continuous spectrum for a spatially flat Universe, and a positive
discrete spectrum for a closed Universe.  The convergence of these
different approaches suggests that the exclusion of a negative
cosmological constant may be a feature of vacuum unimodular or
trace-free Einstein quantum cosmology.

It should be noted, however, that the exclusion of a negative
cosmological constant depends on the formulations and the choice of
matter content.  Once matter is included, negative values of $\Lambda$
become possible.  This is explicitly noted in
Refs.~\cite{Amadei2019b,Amadei2023}, where the free-particle analogy
is employed.  In that framework, the inclusion of matter introduces
bound states in the spectrum, which correspond to a negative
cosmological constant.  Additionally, in covariant unimodular
cosmology including dust matter~\cite{Sahota2025,Mukherjee2026}, a
negative cosmological constant is not excluded, and its implications
are discussed from a different perspective.

For a positive cosmological constant, the eigenstates vanish at $p=0$.
The vanishing of the wave function at $p=0$ corresponds to the
so-called DeWitt boundary condition~\cite{DeWitt1967}, while it is not
imposed as an additional assumption.  We would like to emphasize that
this cannot be interpreted as a dynamical resolution of the
cosmological singularity, such as a quantum bounce studied in
Refs.~\cite{Sahota2025,Chiou2010}.  Rather, it should be regarded as a
kinematical consequence of the unimodular constraint.  In the context
of covariant unimodular gravity, Ref.~\cite{Gielen2025a} discusses the
quantum behavior near zero spatial volume and its implications.  It
would be interesting to examine whether a similar analysis applies in
the present framework.

We have further analyzed the oscillatory sector using the WKB
approximation.  In the unimodular framework, the construction of a
wave packet requires a superposition over the eigenvalues
$\hbar\omega$, which correspond to different values of $\Lambda$.  By
constructing a sharply peaked wave packet and imposing the unimodular
condition, we derived a relation \eqref{eq:68} among the relative
fluctuations of the spatial volume $|p|^{3/2}=V_{o}a^{3}$, the lapse
function $N$, and the eigenvalue $\hbar\omega$.  This feature reflects
a structural difference from standard GR, where the lapse function is
a free gauge parameter and the cosmological constant is not related to
quantum superposition.

The relative fluctuation \eqref{eq:72} can be expressed in terms of
$\Lambda$ and the four-volume $V_{4}$.  In the unimodular framework,
the notion of semiclassicality itself is different from GR.  In
standard GR, semiclassicality is characterized by the suppression of
fluctuations of the geometry around a given classical background.  In
contrast, in unimodular gravity, semiclassicality is not determined
solely by fluctuations of geometry, but also by the spread of the wave
packet formed as a superposition over different values of $\Lambda$.
Equation \eqref{eq:72} therefore indicates how the coherence of such a
superposition depends on $\Lambda$ itself.

When the four-volume $V_{4}$ is identified with the Hubble four-volume
of the present Universe, the relative fluctuation takes an extremely
small value as shown in \eqref{eq:73}.  We would like to emphasize
that this result does not constitute a solution to the cosmological
constant problem.  In the present work, the observed value of the
cosmological constant $\Lambda$ is taken as an external input.  We do
not provide an explanation for the smallness of $\Lambda$ and we do
not propose a mechanism that selects its value.  Rather, the result
indicates that, once the observed value is assumed, the relative
fluctuation becomes sufficiently small to provide a justification for
the semiclassical approximation over cosmological scales.  In this
sense, the unimodular framework is consistent with the observed
Universe.

It is worth noting that the relation between semiclassicality and the
cosmological constant has been discussed in the context of covariant
unimodular gravity.  In Ref.~\cite{Afshordi2022}, it was argued that
semiclassicality in the early Universe provides a lower bound on the
cosmological constant.  It would be interesting to clarify whether
such a bound emerges in the original unimodular framework.  From a
complementary perspective, Ref.~\cite{Chiou2010} suggests that, in
loop quantum cosmology based on covariant unimodular gravity,
semiclassicality is closely related to the smallness of the
expectation value of the cosmological constant.  This viewpoint is
consistent with our results.

In a related context, a similar result of the same
order for the fluctuations of the cosmological constant is also
obtained in causal set theory~\cite{Sorkin2007}. In that framework,
fluctuations of the cosmological constant arise from the fundamental
discreteness of spacetime. Although the physical origin is different
from the present work, both frameworks suggest that the smallness of
the cosmological constant may reflect quantum gravitational
effects.

It is important to understand how general the features found in this
work are within unimodular gravity.  The model studied here is highly
restricted.  It remains to be clarified whether the vanishing of the
wave function at zero volume, the exclusion of a negative cosmological
constant, and the relation among relative fluctuations persist when
matter fields or spatial curvature are included.  Finally, it would
also be worthwhile to examine whether these structures can be traced
back to the full theory.  Clarifying whether these results reflect
genuine features of unimodular gravity or are artifacts of symmetry
reduction remains an important direction for future work.

\section*{Acknowledgments}
The author would like to thank M.~Fukuda for discussions and useful
comments.  The author also would like to thank M.~Geiller, S.~Gielen,
M.~Montesinos, and A.~Salvio for kind and useful comments.

\providecommand{\href}[2]{#2}
\begingroup\raggedright\endgroup
\end{document}